\documentclass[a4paper,11pt]{article}
\pdfoutput=1

\usepackage{a4}
\usepackage{jheppub}
\usepackage{graphics}
\usepackage{amssymb,amsmath}
\usepackage{color}
\usepackage{wasysym}
\usepackage[tight]{subfigure}
\usepackage{booktabs}
\usepackage{multirow}
\usepackage{rotating}
\usepackage{float}
\usepackage{slashed}
\usepackage{ulem}

\newcommand{\beq}{\begin{equation}}
\newcommand{\eeq}{\end{equation}}
\newcommand{\beqa}{\begin{eqnarray}}
\newcommand{\eeqa}{\end{eqnarray}}

\title{The Axiverse induced Dark Radiation Problem}
\author[1,2]{\normalsize{Bobby Acharya}}
\author[1]{\normalsize{Chakrit Pongkitivanichkul}}
\affiliation[1]{Theoretical Particle Physics \&
  Cosmology Group, Department of Physics, King's College London, Strand, London, WC2R 2LS, United Kingdom}
\affiliation[2]{The Abdus Salam International Centre for Theoretical Physics, Strada Costiera 11, Trieste, Italy}
\emailAdd{bobby.acharya@kcl.ac.uk}
\emailAdd{chakrit.pongkitivanichkul@kcl.ac.uk}
\abstract{
The string/$M$ theory Axiverse -- a plethora of very light Axion Like Particles (ALPs) with a vast range of masses -- is arguably a generic prediction of string/$M$ theory. String/$M$ theory also tends to predict that the early Universe is dominated by moduli fields. When the heavy moduli decay, before nucleosynthesis,
they produce dark radiation in the form of relativistic ALPs. Generically one estimates that the number of relativistic species grows with the number of
axions in the Axiverse, in contradiction to the observations that $N_{eff} \leq 4$. We explain this problem in detail and suggest some possible solutions to it.
The simplest solution requires that the lightest modulus decays only into its own axion superpartner plus Standard Model particles and this severely constrains
the moduli Kahler potential and mass matrix.}

\begin{document}
\maketitle
\flushbottom
\section{Introduction}
String/$M$ theory is a theoretical framework which predicts that there are six or seven extra dimensions of space.
Even though the energy scale of the extra dimensions might be large, e.g. at the GUT scale, their lowest energy excitations, the moduli fields,
typically have a much smaller mass, such as the supersymmetry breaking scale. These moduli fields appear in the low energy effective description
of physics as scalar fields which couple to matter through higher dimension operators suppressed by the Planck scale, $m_{pl}$. This is natural
since the moduli are, in fact, extra dimensional gravitons.
In the string/$M$ theory description of physics,
the value of {\it all} Standard Model couplings and masses become the vacuum expectation values of the moduli fields. As an example, the
Maxwell term in the Lagrangian, $L = {1 \over 4 e^2} F_{\mu\nu}^2$ becomes of the form  $  {s \over m_{pl}} F_{\mu\nu}^2$ for a particular modulus field $s$.
The expectation values of the moduli fields are interpreted as determining the size and shape of the extra dimensions and the fact that most Standard Model couplings are weak implies that the extra dimensions are moderately large compared to the fundamental string/$M$ theory scale, which further implies we can usually use a
supergravity Lagrangian to describe the low energy physics.

String/$M$ theory also predicts the existence of axion like particles (ALPs) -- these are periodic, pseudo-scalars which arise as zero modes of the antisymmetric
tensor fields present in all low energy descriptions of string/$M$ theory \cite{Deser:1980rz,Svrcek:2006yi}. In fact, in models with low energy supersymmetry, which is what we will assume from here onwards, the axions and moduli usually pair up to form the complex scalar fields which appear in the chiral supermultiplets in four dimensional supergravity theories.
For example, related to the Maxwell term above will be a term like ${a \over f_a} F_{\mu\nu}\tilde{F}^{\mu\nu}$ where $a$ is an axion field and $f_a$ is the ``axion decay constant", which is typically of
order the compactification scale, and often of order $M_{GUT}$. With this logic: for every $U(1)$ and every simple factor in the full gauge group we will have one modulus field and one axion. Similarly, the magnitudes and phases of the entries of the Standard Model Yukawa matrices are resepctively related to
the moduli and axion vevs. Continuing in this fashion and including supersymmetric couplings as well, we see that the full theory could have hundreds, if not more, moduli and axions. 
 
Since the axions arise as zero modes of antisymmetric gauge fields, the number of axions is determined by the number of harmonic antisymmetric tensor fields on the
extra dimensions. This number is a topological invariant of the extra dimensions (a Betti number), 
so if the extra dimensions have a suitably rich topology there will be a large number of axions present in the
spectrum. One then assumes that ``rich topologies" are generic since simple topologies are rare.
This was the argument given in \cite{Arvanitaki:2009fg} leading to the notion of the string/$M$ theory Axiverse. This is clearly consistent with the argument
given above based simply upon couplings. Note that specific examples of string/$M$ theory compactifications could, and do, exist with very few light axions in the
spectrum. In these examples, the topology is relatively simple and, equivalently, the axions you would expect based on the couplings argument have obtained a large mass
through a direct breaking of the axion shift symmetries by the background geometry.

Extremely weakly coupled scalar fields like moduli and axions can have a considerable impact on cosmological dynamics due to the ``vacuum misalignment" mechanism \cite{Preskill:1982cy, Abbott:1982af, Dine:1982ah}. At very early times when the Hubble scale $H$ is above the masses of these particles, the fields are frozen at order one values ($m_{pl}$ for the moduli and
$f_a$ for the axions). Then, as the Universe expands and $H$ decreases, when $H$ becomes of order$ \sim m_s$ or $m_a$, the equation of motion requires the field to start oscillating around the minimum with a frequency of order $m_s$ or $m_a$. Since the corresponding contribution to the energy density will dilute like matter, even if the Universe was radiation dominated prior to this point, a modulus field will quickly dominate the Universe since its energy density is comparable to radiation at the onset of the oscillations. Next, when the Hubble scale reduces to be of order the modulus decay width, $\Gamma_s \sim {m_s^3 \over m_{pl}^2}$, the modulus field decays. This happens during nucleosynthesis for $m_s \sim$ TeV, which is known as the cosmological moduli problem \cite{Banks:1993en, deCarlos:1993jw}. This problem will be avoided if $m_s \geq 30$ TeV; one could also avoid it by assuming that the Hubble scale after
inflation is always smaller than $m_s$ or if there is a late period of inflation which dilutes the moduli fields, however, both of these options require tuning and are presumably not generic. Therefore, we conclude {\it that string/$M$ theory seems to predict that the early Universe prior to nucleosynthesis is matter dominated.} 

The axions also participate in the vacuum misalignment mechanism but there are important differences. The shift symmetries that these fields enjoy protects their masses from perturbative contributions, hence they receive masses only from non-perturbative effects, such as instantons, which at weak coupling are exponentially small. The resulting very small axion masses means that the axion lifetimes are, unlike the moduli, generically extremely long, with lifetimes that can easily be cosmologically relevant.
Hence, there will be a contribution to the energy density in the form of axion fields today, which behaves as cold dark matter. Notice that the decay of the moduli releases a large amount of entropy which dilutes any relics which existed prior to nucleosynthesis, e.g. ten orders of magnitude dilution is typical. This significantly weakens the upper bound on the QCD axion decay constant, compared to radiation dominated Universes, to be of order $10^{15}$ GeV \cite{Lazarides:1990xp,Fox:2004kb,Kaplan:2006vm,Acharya:2010zx}. This effect also significantly dilutes other relics that may have formed previously, such as domain walls, monopoles or thermal WIMPs.

For our present study, the key point is that the moduli can decay into Standard Model particles, into supersymmetric particles as well as {\it axions}. 
Since the axions are so light and the moduli have masses in the tens of TeV regime, axions produced this way will be relativistic with energies of order several TeV. The expansion of the Universe and precision cosmological observables are sensitive to the relative abundance of relativistic particles. This can be captured by the observable called $N_{eff}$ which
is ``the effective number of neutrino species", but is actually sensitive to all forms of relativistic matter, regardless of how such matter couples to the Standard Model. In that sense, $N_{eff}$ provides a very useful probe of additional, ``hidden," sectors beyond the Standard Model. The Standard Model prediction for $N_{eff}$ at the time of recombination
is 3.045, whilst measurements from CMB observations by WMAP 9-year polarisation data \cite{Hinshaw:2012aka}, South Pole Telescope \cite{Hou:2012xq}, Atacama Cosmology Telescope \cite{Sievers:2013ica}, and Planck 2015 \cite{Ade:2015xua} are $N_{eff}=3.84\pm0.40$ (WMAP9), $N_{eff}=3.62\pm0.48$ (SPT), $N_{eff}=2.79\pm0.56$ (ACT), $N_{eff}=3.15\pm0.23$ (Planck2015) respectively. In a sense, this is a surprising result since one might expect $N_{eff}$ to be much, much larger naively. 

So, from the perspective of string/$M$ theory or the idea of “hidden sectors” more generally, the question actually becomes: {\it why is $N_{eff}$ so small?}
For instance, if, as we have already argued, there are large numbers of light axions and the moduli have significant branching ratios into them, why
isn't $N_{eff}$ of order $N$, the number of axions? We will investigate this question in this paper.

There have been a number of interesting prior studies on axionic dark radiation in string theory \cite{Cicoli:2012aq,Higaki:2012ar,Higaki:2013lra,Conlon:2013isa,Angus:2013zfa,Cicoli:2013ana,Cicoli:2014bfa,Angus:2014bia,Hebecker:2014gka,Cicoli:2015bpq,Marsh:2015xka}. These papers consider examples which have very few light axions. Instead, our interest here is to the dependence of $N_{eff}$ on the number of light axions.

\section{The Axiverse Induced Dark Radiation Problem}

We will illustrate the problem by beginning with a simple model and gradually considering more and more general (realistic) cases as we go on. 

The simplest Lagrangian involving a modulus ($s$), an axion ($t$) and a gauge field strength $F_{\mu\nu}$ is arguably of the form:
\beq
{\mathcal{L} \over {m_{pl}^2}} = {c \over s^2} \partial_{\mu} s \partial^{\mu} s + {c \over s^2} \partial_{\mu} t \partial^{\mu} t +  \tilde{c} s F_{\mu\nu} F^{\mu\nu} - \widetilde{m}^2 ({s - s_o})^2
\eeq
where $c$ and $\tilde{c}$ are constants. $s_o$ reflects that $s$ will have a non-zero vacuum expectation value. In our conventions, $s$ and $t$ are dimensionless and $m_{pl}$ is the Planck mass.

This form of the Lagrangian arises in supersymmetric string and $M$ theory models e.g. the universal axio-dilaton Lagrangian or the model independent axion/modulus
multiplet in heterotic string compactifications \cite{Green:1987mn}. From this Lagrangian we can canonically normalise the fields after setting $s$ to its vacuum value and compute the partial decay widths 

\beq
\Gamma(\hat{s} \rightarrow \hat{t} \hat{t}) = \frac{1}{64\pi c} \frac{m^3}{m_{pl}^2}
\eeq
and
\beq
\Gamma(\hat{s} \rightarrow \gamma \gamma) = \frac{1}{64\pi c} \frac{m^3}{m_{pl}^2}
\eeq
where $m = \frac{\widetilde{m} \langle s \rangle}{\sqrt{2c}}$ is the physical moduli mass.

The contribution to dark radiation of axion from moduli decay can be calculated from \cite{Cicoli:2012aq,Higaki:2012ar,Higaki:2013lra}
\beq
\Delta N_{eff} = \frac{43}{7} \frac{\Gamma_{axions}}{\Gamma_{visible}} \left( \frac{g^{*}}{g^{*}_{reheat}} \right)^{1/3} \label{formula}
\eeq
We can take the decay of moduli into two photons as a {\it model} for the decay of the moduli into Standard Model particles, so this calculation gives
\beq
\Delta N_{eff} \sim O(1)
\eeq
since $\Delta N_{eff}$ is given by the ratio of the decay width of the modulus decay into axions versus Standard Model particles. This illustrates the fact that the moduli couple semi-universally to all particles (as one expects, since, after all they are extra dimensional gravitons).

In this paper we are interested in the case when there are a large number, $N$ of axion/moduli multiplets, $(t_i , s_i)$. The previous Lagrangian can then be generalised to

\beq
\mathcal{L} = {m_{pl}^2} \sum_i ({a_i \over s_i^2} \partial_{\mu} s_i \partial^{\mu} s_i + {a_i \over s_i^2} \partial_{\mu} t_i \partial^{\mu} t_i +\\  
\tilde{a_i} s_i F_{\mu\nu} F^{\mu\nu} - \widetilde{m}_i^2 (s_i^2 - \langle s_i \rangle^2))
\eeq

This Lagrangian arises from a supergravity theory containing $N$ chiral superfields with scalar components $z_j = t_j + i s_j$ with Kahler potential $K = -3 \ln V$
where $V = \Pi_i s_i^{a_i} $. This Kahler potential is a typical term which would arise in string/$M$ theory compactifications\footnote{In the next section we will study more concrete string/$M$ theory examples.}. 
Let us now calculate $N_{eff} \equiv N_{SM} + \Delta N_{eff}$. To do this
we need to evaluate the $N^2$ partial decay widths $\Gamma(\hat{s_j} \rightarrow \hat{t_i} \hat{t_i})$ which can readily be calculated to be
\beq
\Gamma(\hat{s_j} \rightarrow \hat{t_i} \hat{t_i}) = \frac{\delta_{ij}}{64\pi}\frac{1}{a_j} \frac{m_j^3}{m_{pl}^2}
\eeq
where $m_j=\frac{\widetilde{m}_j\langle s \rangle}{\sqrt{2 a_j}}$.
On the other hand we also calculate
\beq
\Gamma(\hat{s_j} \rightarrow \gamma \gamma) = \frac{1}{64 \pi} \frac{1}{( \sum_i \widetilde{a}_i \langle s_i\rangle)^2} \frac{\widetilde{a}^2_j \langle s_j\rangle^2}{a_j} \frac{m_j^3}{m_{pl}^2}
\eeq
which results in
\beq
\Delta N_{eff} = \frac{( \sum_i \widetilde{a}_i \langle s_i\rangle)^2}{\widetilde{a}^2_j \langle s_j\rangle^2} = \frac {1}{(16\pi\alpha)^2 \widetilde{a}^2_j \langle s_j\rangle^2}
\eeq
where we used the fact that the sum which appears is related to the coupling constant of the gauge theory and have set the numerical factors in
equation \ref{formula} to one for simplicity. Important points to note about this example are: 

a) {\it due to the diagonal mass and kinetic terms, a given modulus field $\hat{s}_j$ decays only into its axion partners}; 

b) the moduli with the smallest masses will decay last.

When the last (and lightest)  modulus decays it substantially dilutes the energy density of particles produced from previous decays of heavier moduli. Hence, in computing $\Delta N_{eff}$ we are only interested in axions produced from the lightest moduli fields.

Now, {\it in this particular case} $\alpha$ is interpreted as the fine structure constant evaluated when the moduli decay takes place just before BBN, so $16\pi\alpha$ is an order one number, 
{\it independent of $N$}. On the other hand, since $\frac{1}{16\pi\alpha}$ is a sum of the N terms $\widetilde{a}_j \langle s_j \rangle$, if all $N$ terms contribute
similar amounts to the sum, we would have $N_{eff} \sim N^2$ which is our first indication of the {\it Axiverse induced dark radiation problem}. In this particular, very special model, observational consistency requires that the value of $\alpha$ arises only from the modulus $s_j$ and hence that $\Delta N_{eff}$ is order one or smaller. Let us discuss more typical models.

In much more generality, the moduli dependent kinetic terms are not of the form ${a_i \over s_i^2}$; rather they will be given by more complicated functions which are {\it homogeneous of degree minus two}. This is because the moduli Kahler potentials in string/$M$ theory compactifications can be written as logarithms of homogeneous functions of fixed degrees, which implies that their second derivatives are homogeneous of said degree. 
Thus, one has a kinetic mixing matrix $K_{ij}$ whose entries are homogeneous of degree minus two. Before we discuss this most general case, we consider an intermediate, but instructive case:
models in which the kinetic coefficients are
diagonal, but arbitrary functions of degree minus two, $f_i$. This sort of example occurs when the Kahler potential is dominated by a single term, but which could depend on all the moduli.
In this case we have, setting $m_{pl}=1$:

\beq
{\mathcal{L} = f_i (\partial_{\mu} s_i)^2 + f_i (\partial_{\mu} t_i)^2 + \widetilde{a}_i s_i F_{\mu \nu}^2 + \sum_i \widetilde{m}_i^2 s_i^2}
\eeq
Normalising the fields
\beq
s_i = \frac{1}{\sqrt{2\langle f_i \rangle}} \hat{s}_i, \quad t_i = \frac{1}{\sqrt{2\langle f_i \rangle}} \hat{t}_i, \quad A_{\mu} = \frac{1}{2\sqrt{\widetilde{a}_i \langle s_i \rangle}} \hat{A_{\mu}}
\eeq
gives the Lagrangian
\beq
\mathcal{L} = \frac{1}{2}(\partial_{\mu} \hat{s}_i)^2 +  \frac{1}{2}(\partial_{\mu} \hat{t}_i)^2 + \frac{1}{4} \hat{F}_{\mu \nu}^2 \\+ \frac{\langle \partial_j f_i \rangle}{2\sqrt{2}\sqrt{\langle f_j \rangle} \langle f_i \rangle} \hat{s_j} (\partial_{\mu} \hat{t}_i)^2 \\+ \frac{\widetilde{a}_j}{4\sqrt{2\langle f_i \rangle} \widetilde{a}_i \langle s_i \rangle} \hat{s}_j  \hat{F}_{\mu \nu}^2
\eeq
This results in
\beqa
\Gamma (\hat{s}_j \rightarrow \hat{t_i}\hat{t_i}) &=& \frac{1}{256 \pi} \frac{1}{\langle f_j \rangle} \frac{1}{\langle f_i \rangle^2} \langle \partial_j f_i \rangle^2 \frac{m_j^3}{m_{pl}^2}\\
\Gamma (\hat{s}_j \rightarrow \mbox{Axions}) &=& \frac{1}{256 \pi} \left( \sum_{i=1}^N \frac{1}{\langle f_j \rangle} \frac{1}{\langle f_i \rangle^2} \langle \partial_j f_i \rangle^2 \right) \frac{m_j^3}{m_{pl}^2}\nonumber \\
\Gamma (\hat{s}_j \rightarrow \gamma\gamma) &=& \frac{1}{64\pi} \frac{1}{\langle f_j \rangle} \frac{\widetilde{a}_j^2}{\left(\sum_{i=1}^N \widetilde{a}_i \langle s_i \rangle\right)^2} \frac{m_j^3}{m_{pl}^2}\nonumber
\eeqa
where $m_j = \frac{\widetilde{m}_j}{\sqrt{2\langle f_j\rangle}}$.
The key point here is the sum over $N$ terms in the second of the above equations. If the kinetic coefficient $f_i$ depends on $s_j$ then $s_j$ will be able to
decay into $t_i t_i$ and, in the general case we will have $N$ such decays producing light axions, giving
\beq
\boxed{N_{eff} \propto N}
\eeq
The fact that the decay width 
of the lightest moduli into axions is of order $N$ is independent of the moduli couplings to the hidden sector since it only depends on the number of fields.

It is also instructive to illustrate the $N$-dependence
in simple examples as these demonstrate how the Axiverse induced dark radiation problem might be solved.
In the first example we take all of the kinetic coefficients equal and to be given by
\beq
{f_i = \frac{1}{\sum_k s_k^2} = \frac{1}{s_1^2 + \ldots + s_N^2} \equiv \frac{1}{S_{rms}^2}}
\eeq
The decay width of the j-th modulus to decay into the i-th axion is then
\beq
\Gamma (\hat{s}_j \rightarrow \hat{t}_i \hat{t}_i) = \frac{1}{64\pi} \frac{\langle s_j \rangle^2}{\langle S_{rms}^2 \rangle} \frac{m_j^3}{m_{pl}^2}\\
\eeq
which implies that the total decay width of the j-th modulus to decay into axions is a sum of $N$ terms which adds up to
\beq
\Gamma (\hat{s}_j \rightarrow \mbox{axions}) = \frac{N}{64\pi} \frac{\langle s_j \rangle^2}{\langle S_{rms}^2 \rangle} \frac{m_j^3}{m_{pl}^2}
\eeq
By comparison, the decay width into gauge bosons is
\beq
\Gamma (\hat{s}_j \rightarrow \gamma\gamma) = \frac{1}{64\pi} \frac{\widetilde{a}_j^2 \langle S_{rms}^2 \rangle }{\left(\sum_{i=1}^N \widetilde{a}_i \langle s_i \rangle\right)^2} \frac{m_j^3}{m_{pl}^2}
\eeq
which leads to
\beq
\Delta N_{eff} (s_{j}) = N \frac{\langle s_j \rangle^2}{\langle S_{rms}^2 \rangle}  \frac{\left(\sum_{i=1}^N \widetilde{a}_i \langle s_i \rangle\right)^2}{\widetilde{a}_j^2 \langle S_{rms}^2\rangle} = N \frac{\langle s_j \rangle^2}{\langle S_{rms}^2 \rangle}  \frac{1}{(16\pi \alpha)^2\widetilde{a}_j^2 \langle S_{rms}^2\rangle}
\eeq
Clearly, in this example, we can see that if the vev of $S_{rms}$ is sufficiently large in (11d units)
then one can suppress the axion contribution to the dark radiation density.

Finally, let us discuss the most general case.
The following Lagrangian:
\beq
\mathcal{L} = \sum_{i=1}^N \sum_{j=1}^N C_{ij} U_{ik} s_k \partial_{\mu} t_j \partial^{\mu} t_j
\eeq
is the most general Lagrangian coupling moduli fields to axions with two derivatives of the axion fields. Here, $C_{ij}$ arises from diagonalising the Kahler metric $K_{ij}$ and
$U_{ij}$, which we have ignored until now arises from diagonalising the moduli mass matrix. We supplement this Lagrangian
with typical terms coupling the moduli to Standard Model and supersymmetric particles.
The Lagrangian for moduli-gauge boson interactions is
\beq
\mathcal{L} = \sum_{i=1}^N B_i U_{ik} s_k F_{\mu \nu} F^{\mu \nu} 
\eeq
and the Lagrangian for moduli-scalar kinetic interactions is
\beq
\mathcal{L} =  \sum_{i=1}^N D_i U_{ik} s_k D_{\mu} f D^{\mu} f
\eeq
Dropping numerical factors, the decay width of $s_k$, into various channels is:
\beqa
\Gamma_{\mbox{\tiny{axions}}} &=& \sum_{j=1}^N \Gamma (s_k \rightarrow t_j t_j) \nonumber \\
&=& \sum_{j=1}^N \left( \sum_{i=1}^N C_{ij} U_{ik}\right)^2 \frac{m^3_{s_k}}{M_{PL}^2} \nonumber \\
\Gamma_{\mbox{\tiny{gauge particles}}} &=& n_G \left( \sum_{i=1}^N B_i U_{ik}\right)^2 \frac{m^3_{s_k}}{M_{PL}^2} \nonumber \\
\Gamma_{\mbox{\tiny{fermions/sfermions}}} &=& n_f \left( \sum_{i=1}^N D_i U_{ik}\right)^2 \frac{m^3_{s_k}}{M_{PL}^2} \label{eq:gamma}
\eeqa
where $n_G$ and $n_f$ are the numbers of gauge bosons and fermions respectively. 
Even though the most general model has so many parameters, one can see that we expect $N_{eff} \propto N$:

\beq
\langle \Delta N_{eff} \rangle \propto \frac{\Gamma_{axions}}{\Gamma_{visible}} \propto \frac{N \langle C\rangle^2}{n_G \langle B\rangle^2+n_f \langle D\rangle^2} \propto N
\eeq

This arises because we expect the mean values $C$, $B$ and $D$ to be comparable and that $(\sum_{i=1}^N U_{ik})^2$ to be order one. This is borne out by explicit calculations, see e.g. \cite{Acharya:2007rc,Acharya:2008bk,Acharya:2008zi,Acharya:2008hi,Balasubramanian:2005zx,Conlon:2005ki,Cicoli:2012aq,Higaki:2012ar,Higaki:2013lra}.
In other words, since the moduli couplings to axions are comparable to their couplings to the Standard Model particles, the string/$M$ theory axiverse is in serious tension with observed limits on the amount of dark radiation.
In special examples with low numbers of axions, one can see that it is possible to generate acceptably small amounts of dark radiation assuming certain couplings are small enough, for example, \cite{Cicoli:2012aq,Higaki:2012ar,Higaki:2013lra,Conlon:2013isa,Angus:2013zfa,Angus:2014bia,Hebecker:2014gka,Cicoli:2015bpq}. But in general, this will be difficult to avoid.

\section{String/$M$ theory examples}

\subsection{Calabi-Yau Compactifications}
In Calabi-Yau compactifications of superstring theories to four dimensions, The moduli and axion kinetic terms in the Lagrangian are derived from a function of the moduli fields called the Kahler potential, $K$, which, up to a coefficient is given by
\beq
K = -a \ln V_X
\eeq

Here, $V_X$ is the volume of the Calabi-Yau manifold (as a function of the moduli). This is a sum of
terms with coefficients given by the triple intersection numbers $d_{ijk}$. The coefficient $a$ takes different values, depending upon which string theory one is considering.
In the heterotic and Type IIA compactifications, the volume is given as a function of the Kahler moduli $S_i$ as:

\beq
V_X = \sum_{i=1}^n d_{ijk} S_i S_j S_k
\eeq

Clearly, in a completely generic case, with many non-zero entries in $d_{ijk}$,
$V_X$ is a sum of many terms and $K_{ij}$ will not be diagonal. Hence, upon diagonalisation,
when expanding around a particular vacuum state, the matrices $C_{ij}$, $U_{ij}$ and the coefficients $B_i$ and $D_i$ will be quite general and we expect $N_{eff} \propto N$.


In the LARGE volume scenario of \cite{Balasubramanian:2005zx,Conlon:2005ki}, there is a modulus field with a vev much larger than that of the other moduli.
In this case, the
volume functional of the Calabi-Yau threefold is approximated by
\beqa
V &=& s_1^{3/2}-s_2^{3/2}-\ldots-s_N^{3/2} \\
K &=& -2 \ln{V}
\eeqa
In the limit where the $s_1$ vev is larger than the other vevs, $s_1 \gg s_i$, the diagonalised Kahler metric is approximately
\beq
f_1 = K_{11} \approx \frac{3}{4 s_1^2},\quad f_i = K_{ii} \approx \frac{3}{8 s_1^{3/2} s_i^{1/2}}
\eeq
\beq
\partial_1 f_1 = -\frac{3}{2 s_1^3},\quad \partial_i f_1 = 0,\quad \partial_1 f_i = -\frac{9}{16 s_1^{5/2} s_i^{1/2}}, \quad \partial_i f_i = - \frac{3}{16 s_1^{3/2} s_i^{3/2}}
\eeq
For $s_1$, it turns out that $\Delta N_{eff} \propto N$. This can be seen as follows. The decay widths to axions are
\beqa
\Gamma (\hat{s}_1 \rightarrow \hat{t}_1 \hat{t}_1) &=& \frac{1}{256\pi} \left( \frac{16}{3} \right) \frac{m^3}{m_{pl}^2}\\
\Gamma (\hat{s}_1 \rightarrow \hat{t}_{i\neq1} \hat{t}_{i\neq1}) &=& \frac{3}{256\pi} \frac{m^3}{m_{pl}^2}\\
\Gamma (\hat{s}_1 \rightarrow \mbox{axions}) &=& \frac{1}{256\pi}\left(\frac{16}{3} + 3 (N-1)\right) \frac{m^3}{m_{pl}^2}
\eeqa
whilst the gauge boson channel gives
\beq
\Gamma (\hat{s}_1 \rightarrow \gamma\gamma) = \frac{1}{48\pi} \frac{\widetilde{a}_j^2 \langle s_1^2 \rangle }{\left(\sum_{i=1}^N \widetilde{a}_i \langle s_i \rangle\right)^2} \frac{m^3}{m_{pl}^2}
\eeq
resulting in a dark radiation contribution of
\beq
\Delta N_{eff} (s_{1}) = \left(1+\frac{9}{16}(N-1)\right)\frac{\left(\sum_{i=1}^N \widetilde{a}_i \langle s_i \rangle\right)^2}{\widetilde{a}_1^2 \langle s_1^2\rangle}
\eeq

This is interesting, because in LARGE volume models, the vev $s_1$ is expected to be much larger than the other vevs, hence one expects a suppression of $\Delta N_{eff}$ in this case, following our discussion in section two. Furthermore, $s_1$ is typically the lightest modulus in this scenario \cite{Conlon:2005ki}.

For completeness, for $s_{j\neq1}$, the dark radiation density doesn't depend on $N$ :
\beqa
\Gamma (\hat{s}_{j\neq1} \rightarrow \hat{t}_{j\neq1} \hat{t}_{j\neq1}) &=& \frac{1}{128\pi} \frac{ \langle s_1 \rangle^{3/2}}{\langle s_j \rangle^{3/2}} \frac{m^3}{m_{pl}^2} \\
\Gamma (\hat{s}_{j\neq1} \rightarrow \hat{t}_{j\neq i} \hat{t}_{j\neq i}) &=& 0 \\
\Gamma (\hat{s}_{j\neq1} \rightarrow \mbox{axions}) &=& \frac{1}{128\pi} \frac{ \langle s_1 \rangle^{3/2}}{\langle s_j \rangle^{3/2}} \frac{m^3}{m_{pl}^2}
\eeqa
gauge boson channel is
\beq
\Gamma (\hat{s}_{j\neq1} \rightarrow \gamma\gamma) = \frac{1}{24\pi} \frac{\widetilde{a}_j^2 \langle s_j^{1/2} \rangle \langle s_1^{3/2} \rangle }{\left(\sum_{i=1}^N \widetilde{a}_i \langle s_i \rangle\right)^2} \frac{m^3}{m_{pl}^2}
\eeq
So the total dark radiation density is proportional to
\beq
\Delta N_{eff} (s_{j\neq1}) =\frac{3}{16}\frac{\left(\sum_{i=1}^N \widetilde{a}_i \langle s_i \rangle\right)^2}{\widetilde{a}_j^2 \langle s_j^2\rangle}
\eeq

\subsection{Diagonal Kahler metrics}

Clearly, from the above discussions, one can suppress dark radiation from moduli decays when the Kahler metric for the moduli fields is approximately diagonal. This will be the case when the Volume function is dominated by just one term only.

\beq
V_X = \prod_{i=1}^N S_i^{a_i},\quad K = -3 \sum_{i=1}^N a_i \ln S_i \label{eq:G2Kahler}
\eeq
where $a_i$ are microscopic parameters whose sum is a constant determined by the geometry of the extra dimensions. This is unity for the Calabi-Yau case and $\frac{7}{3}$ for $G_2$-manifolds. To calculate decay width, let us translate the above internal manifold into condition on decay width coefficients particularly $C_{ij}$. It can be shown that (see appendix A)
\beqa
C_{ij} &=& \frac{1}{\sqrt{K_{ii}^D}} \frac{\partial \ln K_{ii}^D}{\partial s_j} \nonumber \\
B_{i} &=& \frac{\alpha}{\sqrt{K_{ii}^D}} N_i\\
D_{i} &=& \frac{1}{\sqrt{K^D_{ii}}} \frac{\partial \ln K_{\alpha \alpha}^D}{\partial s_i} \nonumber
\eeqa
where $K^D_{ij}$ is the diagonal Kahler metric. From (\ref{eq:G2Kahler}), It is trivial to show that the coefficients are also diagonal:
\beq
C_{ij} = C_i\delta_{ij} \label{eq:cijdiag}
\eeq
Following the previous analysis, this simple relation implies that $\Delta N_{eff}$ becomes independent of $N$ on average.
\beq
\langle \Delta N_{eff} \rangle \propto \frac{\langle C\rangle^2}{n_G \langle B\rangle^2+n_f \langle D\rangle^2}
\eeq
where the orthogonality of rotation matrix, $(\sum_{i=1}^N U_{ik}^2) = 1$, has been used.

The physical reason for this behaviour is that this particular volume form forces each modulus to decay only into its axionic partner. If we assume further that this basis is already physical, i.e. there is no further mixing between moduli or axions, it becomes clear that dark radiation, regardless of N, consists of only one species of axion which is the partner of the last modulus to decay. 

The moduli mixing matrix can also play a role in suppressing dark radiation.
Again, though non-generic, this occurs when there is a relation between the moduli mass matrix and the eigenvalues of the Kahler metric: 
\beq
\sqrt{K^D_{ii}}\propto U_{ij} \label{eq:sup}
\eeq
The above relation is equivalent to 
\beq
\frac{1}{C_i} \propto U_{ik}, \; \frac{1}{B_i} \propto U_{ik}, \;\frac{1}{D_i} \propto U_{ik}
\eeq
In this case, the correlation becomes
\beq
\langle \Delta N_{eff} \rangle \propto \frac{N\langle C\rangle^2}{n_G N^2 \langle B\rangle^2 +n_f N^2\langle D\rangle^2} \propto \frac{1}{N}
\eeq
Therefore, under these very special circumstances, dark radiation can actually be suppressed by the number of axions on average. This counter-intuitive result is merely the effect of increasing N-dependence of the moduli to visible sector couplings so that dark radiation is dominated by standard model radiation (neutrinos). Most likely, this is merely a curious observation rather than a realistic case.

\subsubsection{Mass matrix in $G_2$ compactified M-theory}

In this subsection, we put together some of these results in a concrete setting where the moduli mass matrix is known, namely $G_2$-compactified $M$ theory. 
Again, we are assuming that $K$ is dominated by a single term:
\beq
K=-3\ln\left(\prod_{i=1}^N s_i^{a_i}\right)
\eeq
where $\sum_{i=1}^N a_i = \frac{7}{3}$. From above, $N_{eff}$ becomes independent of the number of axions in this model. 
However, regardless of this advantage, one could easily find that
the typical value of $N_{eff}$, although independent of $N$, is actually too large in practice e.g. $\Delta N_{eff} \sim 10$. 
We would therefore like to investigate the possibility of further suppressing dark radiation in this setup.

We briefly recall some details of moduli stabilisation. It has been shown in \cite{Acharya:2007rc} that with a hidden sector with two gauge groups where first group is sQCD with 1 flavour of quarks and second group is pure glue sQCD leads to dS vacua. The superpotential is written as
\beq
W = A_1 \phi^a e^{ib_1\sum^N_i N_i S_i} + A_2 e^{ib_2 \sum^N_i N_i S_i}
\eeq
where $\phi$ is the meson superfield in the hidden sector.
With the superpotential and Kahler potential being specified, it is straightforward yet tedious to work out the mass mixing matrix resulting from moduli stabilisation \cite{Acharya:2008hi,Acharya:2008bk}.
\beqa
U_{kj} = \sqrt{\frac{a_{j+1}}{(\sum_{i=1}^{j} a_i)(\sum_{i=1}^{j+1} a_i)}}\sqrt{a_k}, &\mbox{, }& k \leq j\nonumber \\
U_{kj} = -\sqrt{\frac{\sum_{i=1}^{j} a_i}{\sum_{i=1}^{j+1} a_i}} &\mbox{, }& k = j+1\nonumber\\
U_{kN} = \sqrt{\frac{3 a_k}{7}} & & \label{eq:rotmat}
\eeqa
where $i=1\ldots N-1$ are the degenerate light moduli and $i=N$ is heavy modulus. Notice that except $k=j+1$, $U_{kj}\propto \sqrt{a_k} \propto \sqrt{K_k}$. Therefore, we the element $U_{j+1,j}$ will be suppressed if it turned out that:
\beq
\sum_{i=1}^{j} a_i \ll a_{j+1} \label{eq:G2suppress}
\eeq
As a result, one would expect $\frac{1}{N}$ suppression on dark radiation under this condition.

The modulus decay width can be calculated from \cite{Acharya:2008bk}
\beqa
\Gamma_{X_j} &=& D_{X_j} \frac{m_{X_j}^3}{M_{Pl}^2} \nonumber \\
D_{X_j} &=& \alpha \left( \sum_{k=1}^N \frac{U_{kj}^2}{a_k}\right) + \beta \left( \sum_{k=1}^N \frac{U_{kj}}{\sqrt{a_k}} \right)^2 \label{eq:decaytot}
\eeqa
where $\alpha$ and $\beta$ are index-independent parameters dependent on the microscopic details of the $G_2$ manifold. The first term represents the decay width into axions where the latter represents decay width into visible particles. 
From (\ref{eq:rotmat}) and (\ref{eq:decaytot}) it is trivial to see that total decay width of $j^{th}$ modulus and corresponding dark radiation are controlled by 
\beqa
\Gamma_j &\propto& \frac{\sum_{i=1}^{j} a_i}{a_{j+1}} \\
\Delta N_{eff}(X_j) &\propto& \frac{j a_{j+1}^2 + (\sum_{i=1}^{j} a_i)^2}{(j a_{j+1} - \sum_{i=1}^{j} a_i)^2}
\eeqa
Applying (\ref{eq:G2suppress}), we clearly see that $\Delta N_{eff}(X_j) \propto \frac{1}{j}$ and $\Gamma_j$ becomes smallest. This is essential to the model because it guarantees that the last decay modulus exhibits $\frac{1}{N}$ behaviour. For practical purpose, only $j = N-1$ in condition (\ref{eq:G2suppress}) will be assumed.

Next, we will explicitly show correlations between the number of axions and $N_{eff}$. Instead of scanning the $N$ parameters $a_i$ space, we will give systematic examples of simple configurations of the $a_i$ which work:

The first example is when $n$ of the $a_i$ are large and the rest small:
\beq
a_i = \{ \underbrace{\epsilon \bar{a}, \ldots, \epsilon \bar{a}}_{N-n}, \underbrace{\frac{7}{3 n} - \frac{\epsilon \bar{a} (N-n)}{n}, \ldots, \frac{7}{3 n} - \frac{\epsilon \bar{a} (N-n)}{n}}_n \}
\eeq

The second is a “geometric sequence” of $a_i$’s,
\beq
a = \{ a_0, a_0, a_0 r, a_0 r^2, \ldots, a_0 r^{N-2}\}
\eeq

Though these can be viewed as toy models at best, they both illustrate that, in principle,
the amount of dark radiation can actually {\it decrease} as one increases the number of axions. This is illustrated in the two figures.

\begin{figure}[t]
\centering
\subfigure[]{\includegraphics[width=0.48\textwidth]{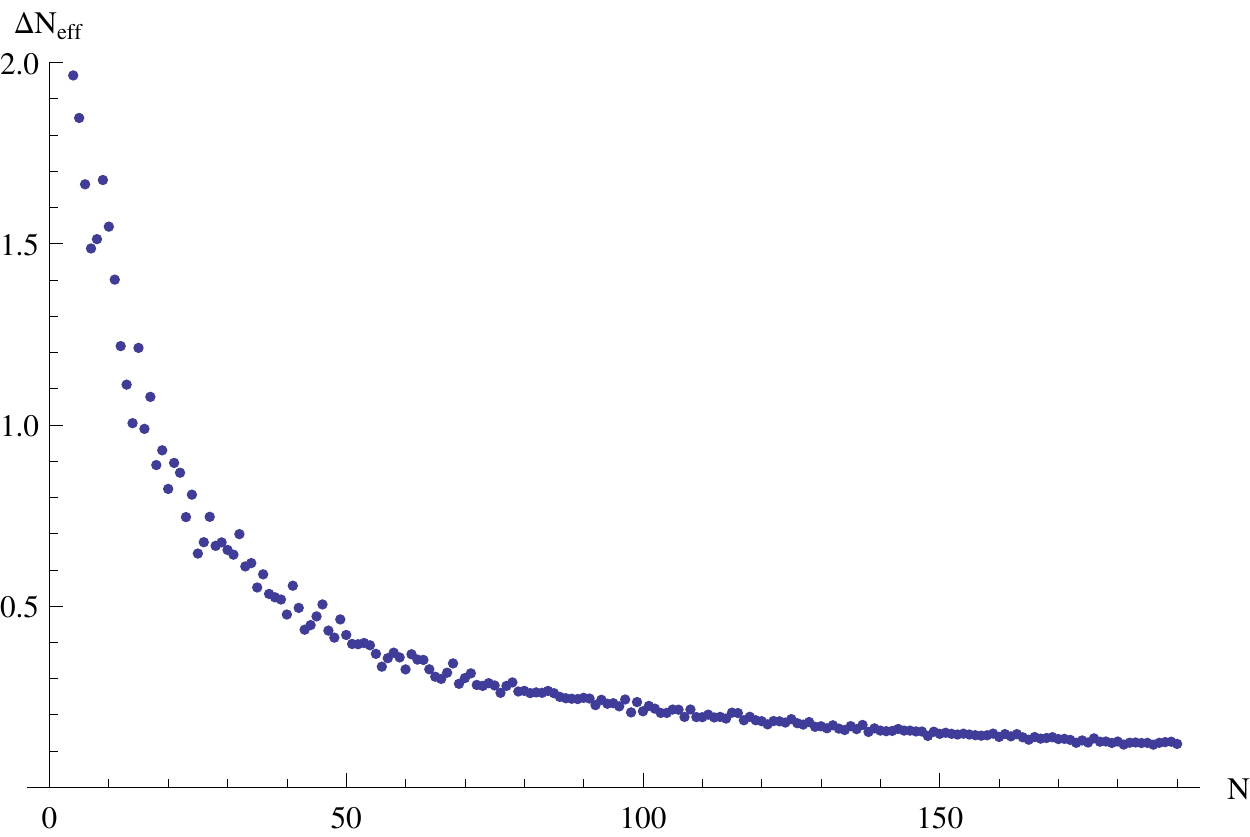}}
\subfigure[]{\includegraphics[width=0.48\textwidth]{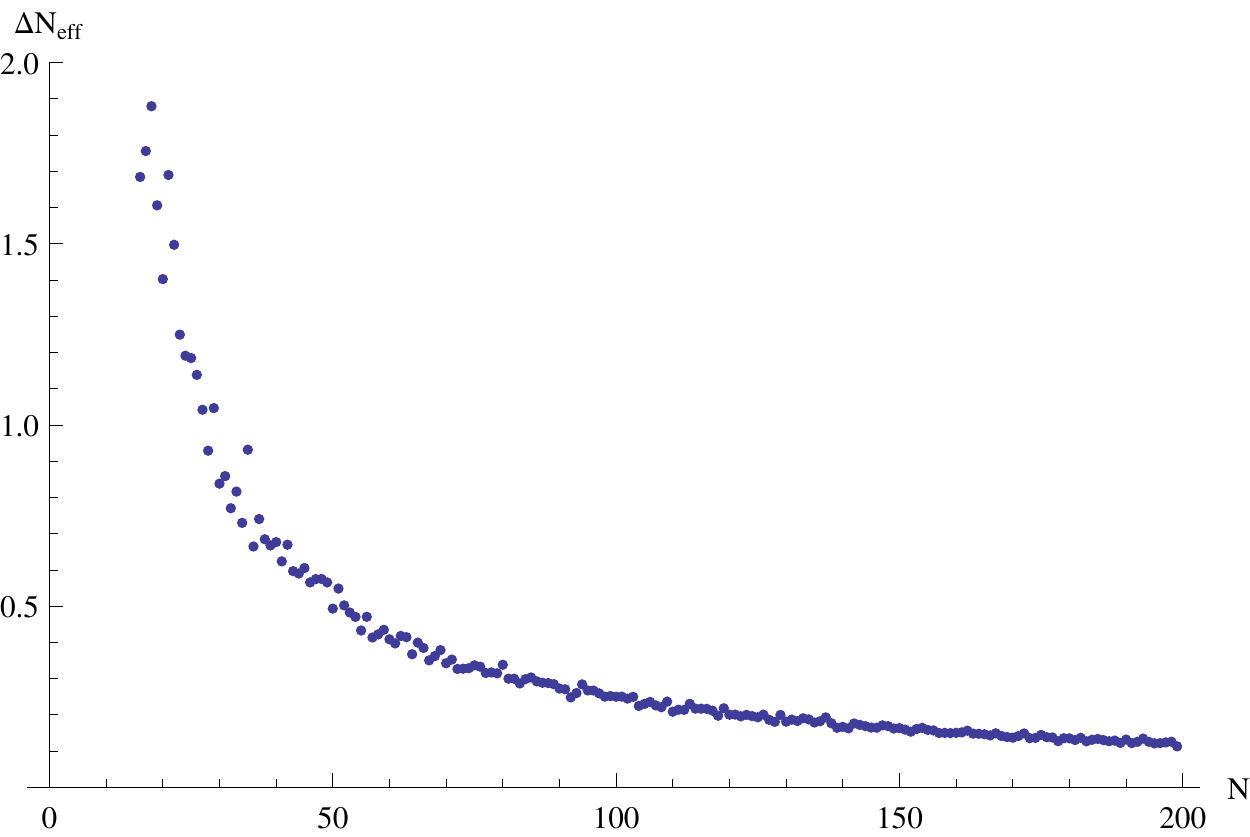}}
\caption{Left: Result from geometric sequence configurations showing $\Delta N_{eff}$ as a function of $N$, where $r = 2$. Right: Result from double moduli dominated configurations showing $\Delta N_{eff}$ as a function of $N$, where $\epsilon N = 0.1$ \label{MGS1N}}
\end{figure}

\begin{figure}[t]
\centering
\subfigure[]{\includegraphics[width=0.48\textwidth]{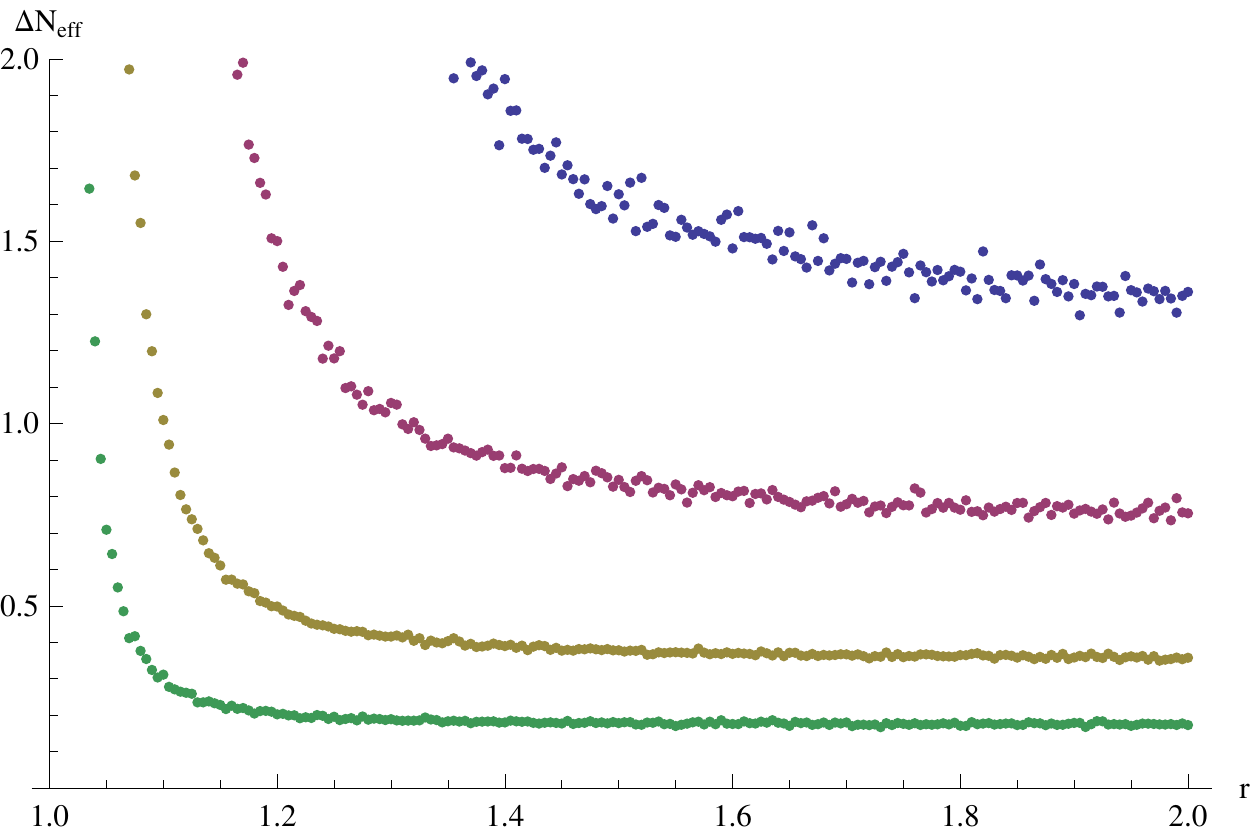}}
\subfigure[]{\includegraphics[width=0.48\textwidth]{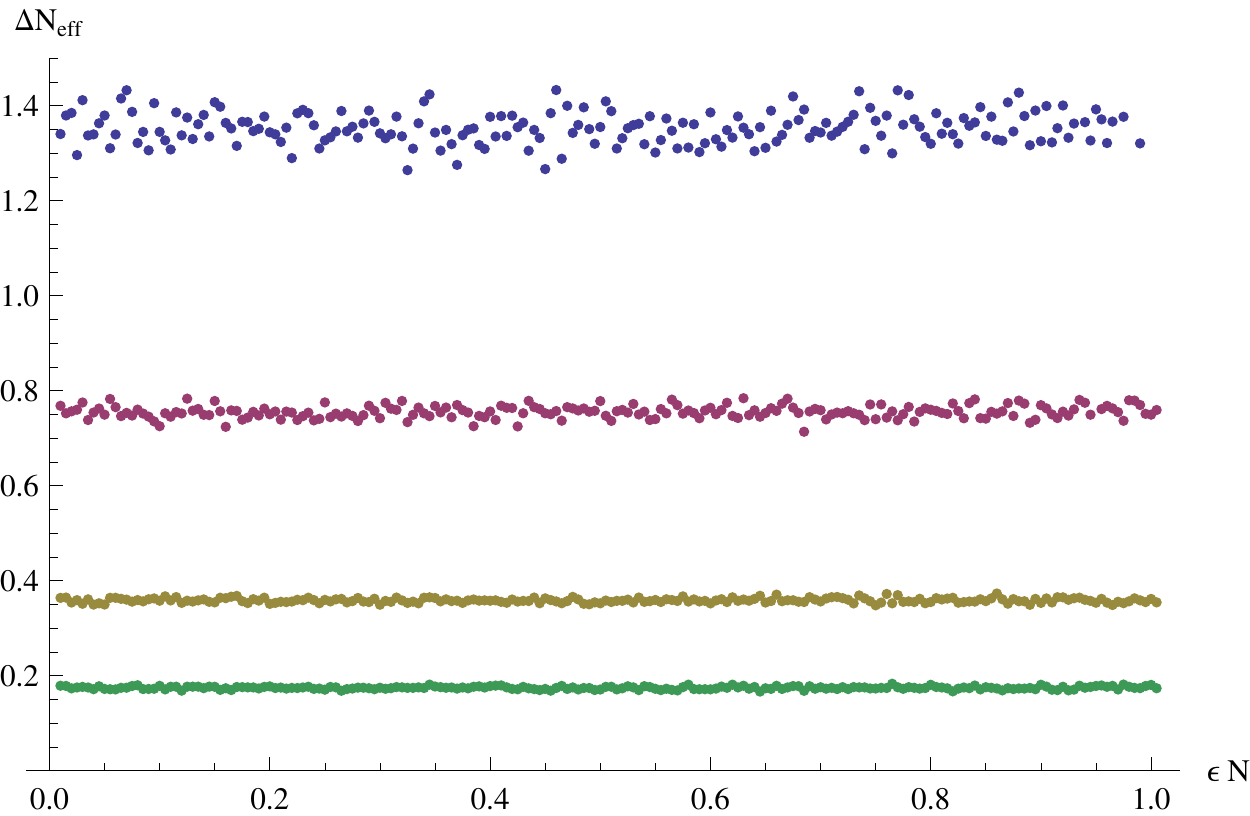}}
\caption{Left: Result from geometric sequence configurations showing $\Delta N_{eff}$ as a function of $r$, where points in blue, red, yellow, green are $N = 30, 50, 100, 200$ respectively. Right: Result from double moduli dominated configurations showing $\Delta N_{eff}$ as a function of $\epsilon N$, where points in blue, red, yellow, green are $N = 30, 50, 100, 200$ respectively \label{MGS}}
\end{figure}



\section{Conclusions and Outlook}
$\Delta N_{eff}$ is a very powerful probe of light degrees of freedom in the hidden sector and, somewhat surprisingly, has been constrained to be quite small, consistent with zero. 
The Axiverse induced Dark Radiation Problem arises from the plethora of light degrees of freedom that can be present in string/$M$ theory compactifications to four dimensions. Though we focused on the axions, similar conclusions can be drawn from hidden photons and other light particles in the hidden sector. We pointed out several possible mechanisms via which this problem could be avoided: a) a relatively large modulus vev as in the LARGE volume scenario; b) alignment between the axion kinetic and mass mixing matrices so that the last modulus to decay does so predominantly into its axionic partner. It would be very interesting to explore these mechanisms in more detail in various specific models. One potential problem with the large vev solution in practice is that the large vev corresponds to a weak Standard Model coupling. In general, it might be difficult to make the vev large enough without making the Standard Model coupling too small.

\acknowledgments
The work of BSA is supported by the UK STFC via the research grant ST/J002798/1. CP is supported by the KCL NMS graduate school and ICTP Trieste.

\appendix
\section{Decay coefficients}
We show detail analysis for coefficients $C_{ij}$, $B_i$ and $D_i$ in this section. The kinetics terms for moduli and axions are controlled by Kahler metric as following
\beq
\mathcal{L} = \frac{1}{2} K_{ij} \partial^{\mu} s^i \partial_{\mu}s^j + \frac{1}{2} K_{ij} \partial^{\mu} t^i \partial_{\mu} t^j
\eeq
After canonically normalisation of moduli and axions, we can expand Kahler metric as a function of moduli field. After taking moduli mixing into account, the result for interaction Lagrangian is
\beq
\mathcal{L}_{\tilde{s}_k \tilde{t}_i \tilde{t}_i} = \frac{1}{2}\frac{\sum_{j=1}^N \frac{\partial K_{ii}^D}{\partial s_j}U_{jk}}{(K_{ii}^D)^{3/2}} \tilde{s}_k  \partial^{\mu} \tilde{t}^i \partial_{\mu} \tilde{t}^j
\eeq
where $K^D$ is Kahler metric after diagonalisation. $\tilde{s}$, $\tilde{t}$ are canonically normalised fields after mixing. Straightforwardly, one can derive decay width into axions as
\beq
\Gamma_{\mbox{axions}} = \frac{1}{32\pi}\sum_{i=1}^N \left( \sum_{j=1}^N \frac{1}{\sqrt{K_{ii}^D}} \frac{\partial \ln K_{ii}^D}{\partial s_j}U_{jk} \right)^2 \frac{m_{X_k}^3}{M_{pl}^2} \label{eq:axwidth}
\eeq
For gauge sector, the Lagrangian takes the form
\beq
\mathcal{L} = -\frac{1}{4} \left( \sum_{i=1}^N N_i z_i \right) F_{\mu \nu} F^{\mu \nu}
\eeq
After canonically normalisation of moduli and gauge fields and mixing between moduli, we get the interaction terms between moduli and gauge fields.
\beq
\mathcal{L} = \frac{1}{4}\frac{1}{\left(\sum_{i=1}^N N_i \langle s_i \rangle\right)} \sum_{i=1}^N \frac{N_i U_{ik}}{\sqrt{K_{ii}^D}} \tilde{s}_k F_{\mu \nu} F^{\mu \nu}
\eeq
The moduli decay width into gauge bosons/gauginos is given by
\beq
\Gamma_{\mbox{gauge}} = \frac{N_G}{32 \pi} \left( \sum_{i=1}^N \frac{\alpha}{\sqrt{K_{ii}^D}} N_i U_{ik} \right)^2 \frac{m_{X_k}^3}{M_{pl}^2} \label{eq:gwidth}
\eeq
For matter sector, the interaction terms can be found from
\beq
\mathcal{L} = K_{\alpha \beta} D_{\mu} f^{\alpha} D^{\mu} f^{\beta} + K_{\alpha \beta} \tilde{f}^{\alpha} \slashed{D} \tilde{f}^{\beta}
\eeq
Then, after normalisation of moduli field and fermions/sfermions fields, the interaction terms become
\beq
\mathcal{L} = \sum_{i=1}^N \frac{1}{\sqrt{K^D_{ii}}} \frac{\partial \ln K_{\alpha \alpha}^D}{\partial s_i}U_{ik} \tilde{s}_k \left( D_{\mu} f^{\alpha} D^{\mu} f^{\alpha} + \tilde{f}^{\alpha} \slashed{D} \tilde{f}^{\alpha} \right)
\eeq
Therefore, the decay width into fermion can be written as
\beq
\Gamma_{\mbox{fermion}} \propto \left(  \sum_{i=1}^N \frac{1}{\sqrt{K^D_{ii}}} \frac{\partial \ln K_{\alpha \alpha}^D}{\partial s_i}U_{ik}  \right)^2 \frac{m_{X_k}^3}{M_{pl}^2} \label{eq:fwidth}
\eeq
Comparing to previous section, we get
\beqa
C_{ij} &=& \frac{1}{\sqrt{K_{ii}^D}} \frac{\partial \ln K_{ii}^D}{\partial s_j} \nonumber \\
B_{i} &=& \frac{\alpha}{\sqrt{K_{ii}^D}} N_i \label{eq:coeffdetail}\\
D_{i} &=& \frac{1}{\sqrt{K^D_{ii}}} \frac{\partial \ln K_{\alpha \alpha}^D}{\partial s_i} \nonumber
\eeqa


\bibliographystyle{JHEP}

\bibliography{dneffG2ref}

\end{document}